\documentclass[reqno,12pt,draft]{article}
\usepackage{amsmath,amssymb,amsfonts,epsfig,graphicx,euscript}

\newcommand{\bse}{\begin{subequations}}
\newcommand{\ese}{\end{subequations}}
\newcommand{\be}{\begin{equation}}
\newcommand{\ee}{\end{equation}}
\newcommand{\bea}{\begin{eqnarray}}
\newcommand{\eea}{\end{eqnarray}}
\newcommand{\ba}{\begin{array}}
\newcommand{\ea}{\end{array}}
\def\dm{{\dot{\mu}}}
\def\dn{{\dot{\nu}}}
\def\da{{\dot{\alpha}}}
\def\db{{\dot{\beta}}}

\begin{document}

\begin{flushright}
\end{flushright}
\begin{flushright}
\end{flushright}
\hfill%
\vbox{
    \halign{#\hfil        \cr
           IPM/P-2010/002\cr
                     }
      }
\vspace{1cm}
\begin{center}

{\LARGE {\sc Multiple D$_p$-branes as a D$_{p+2}$-brane \\}}

\bigskip
{ M. Ali-Akbari\footnote{aliakbari@theory.ipm.ac.ir}}\\
{School of Physics,\\ Institute for Research in Fundamental Sciences(IPM),\\
P. O. Box 19395-5531, Tehran, Iran }
\\
\bigskip
and
\bigskip
\\
{M. A. Ganjali\footnote{ganjali@theory.ipm.ac.ir}} \\
{Department of Fundamental Sciences,\\ Tarbiat Moaallem
University,\\P. O. Box 31979-37551, Tehran, Iran}

\end{center}

\bigskip
\begin{center}
{\bf { Abstract}}\\
\end{center}
From BFSS matrix theory considerations,
it is expected that a single D$_{p+2}$-brane action can be obtained from N D$_p$-brane action in large
N limit. We examine and confirm this expectation by working out the details
of DBI and Chern-Simons terms of D$_{p+2}$-brane
action from D$_{p}$-brane action. We show that the same relation works for non-BPS, as well as BPS branes.

\newpage
\section{Introduction}
D$_p$-brane is a $(p+1)$-dimensional hypersurface in space-time
defined by the property that open strings can end on it. They have
been an active area of study and many remarkable properties of
them have been discussed \cite{Polchinski}. In
particular, they have provided a useful tool for the study of black
holes in string theory \cite{Strominger}; AdS/CFT correspondence
is another offspring of D$_p$-branes \cite{magoo}. Moreover, it is
well-known that IIA and IIB string theories have odd and even
BPS D$_p$-branes, respectively, and T-duality changes D$_p$-branes in
these theories to each other (or in other words it replaces a
scalar field with a gauge field and vice versa). D$_p$-branes are
sources of RR (P+1)-form field in IIA and IIB string theories.

There is an interesting candidate for DLCQ of M-theory
in terms of D$_{0}$-branes, the BFSS matrix model
\cite{Banks}. This conjecture tells us that all D$_p$-branes in IIA
superstring theory can be described in terms of D$_0$-branes and
their bound states. This is a requirement for the BFSS matrix model
as it is supposed that
the theory of D$_0$-branes describes the whole theory of DLCQ of
M-theory. In particular, e.g. a flat D$_2$-brane may be understand as
a bound state of N(N$\rightarrow\infty$) D$_0$-branes.
Therefore, understanding of connection among various
D$_p$-branes seems to be significant. It is well known that the low energy limit
of the D-brane action reduce to the Yang-Mills(YM) action. In YM limit, This
connection has already been addressed in the literature e.g. see
\cite{Susskind:2001fb} but here we use another way introduced in \cite{Ho} and
explain it more precisely. In fact in YM theory we consider
fluctuations around two directions which are transverse to multiple
D$_p$-branes and it will be shown that these two directions play the
role of components of gauge field in the world volume of
D$_{p+2}$-brane. We will then extend the prescription found at the
level of YM theory to terms, for both BPS and non-BPS
Dirac-Born-Infeld(DBI) action and their Chern-Simons actions as
well. The D$_{p+2}$-brane action is consistent with what is
expected.

The multiple D$_p$-branes theory has three parameters the string
length, $l_s$, string coupling constant, $g_s$, and the number of
D$_p$-branes, $N$. Then our prescription can be explained as follow.
Two transverse directions can be considered as a \emph{fuzzy}
torus\footnote{Although, other configurations can be considered.}
(compact transverse direction on fuzzy torus) or on the other
hand, for finite $N$, D$_p$-branes are uniformly distributed on these two
directions. Now two new parameters are added to our theory which are
radii of fuzzy torus, $R_1,R_2$. In order to take geometric
interpretation, there is a consistent limit which is when the number
of multiple D$_p$-branes and radii of fuzzy torus go to infinity
like $\theta=\frac{R_1R_2}{l_s^2N}\rightarrow 0$. As we will see in
this limit D$_p$-branes have been dissolved in the world volume of a
D$_{p+2}$-brane.\footnote{The other limits are when $\theta$ is constant or
goes to infinity. In both cases the final theories are non-commutative.}

In comparison to two T-dualities replacing two transverse directions with two
components of gauge field(or vice versa) on D$_p$-branes, in our
prescription, we consider \emph{multiple} D$_p$-branes and show
that two transverse directions replace with two components of gauge
filed on D$_{p+2}$-brane. In opposite way, one can argue that a
D$_{p+2}$-brane becomes \emph{multiple} D$_p$-branes if two
components of the gauge field are substituted by transverse scalar
fields.

This paper is organized as following.
In the two next sections the prescription is introduced and applied to YM theory
and DBI actions. We then show that the non-BPS D-branes
are consistent with above prescription and in the section five Chern-Simons action
is considered. The last section is devoted to conclusion.

\section{A large N limit of SYM theory}
The bosonic part of a $p+1$ dimensional U(N) SYM theory is described by the action %
\be\label{action} %
 S=-\frac{1}{2g_{YM}^2}\int d^{p+1}x{\rm{Tr}}\left((D_\mu
 X^I)^2-\frac{1}{2}[X^I,X^J]^2+\frac12F_{\mu\nu}^2\right),
\ee %
where\footnote{In terms of D-brane and string theory parameters note that
$g_{YM}^2=(\lambda^2T_p)^{-1}=(2\pi\lambda)^{\frac{p-1}{2}}\lambda^{-1}g_s,
\ \ \lambda=2\pi l_s^2$ where $T_p$ is the brane tension.}  %
\be\begin{split} %
 F_{\mu\nu}&=\partial_\mu A_\nu-\partial_\nu A_\mu+i[A_\mu,A_\nu],\cr
 D_\mu\Phi&=\partial_\mu\Phi+i[A_\mu,\Phi].
\end{split}\ee %
$\mu(=0,...,p)$ and $I(=p+1,..,9)$. This action is also the action for
N coincident D$_p$-branes in the $\alpha'\rightarrow 0$ limit and in this picture
$\mu$ and $I$ denote world volume and transverse indices respectively.
The above action has following U(N) gauge symmetry %
\bse\begin{align} %
 \label{xtransformatiom}\delta_gX^I&=i[X^I,\alpha],\\
 \delta_gA_\mu&=\partial_\mu\alpha+i[A_{\mu},\alpha].
\end{align}\ese %
In order to describe a D$_{p+2}$-brane, our prescription two steps. %
\begin{itemize}
    \item
    The $9-p$ transverses directions are decomposed as
    \be %
     X^I\sim(y^{\dm=1,2},X^{i=p+3,..,9}),
    \ee %
    and the following replacement will then be done
    \bse\label{prescription}\begin{align} %
     \label{prescription1}[\Phi(x),\Psi(x)]&\rightarrow
     i\lambda\theta\{\Phi(x,y),\Psi(x,y)\},\\
     \label{prescription2}\int{d^{p+1}x \rm{Tr}(\star) }&\rightarrow
     \frac{1}{\lambda}\int{d^{p+1}xd^2y(\star)},
    \end{align}\ese%
    where $\{\Phi,\Psi\}=\epsilon^{\dm\dn}\partial_\dm \Phi\partial_\dn\Psi$(it is fixed such that
    $\epsilon_{\dm\dn}\{y^\dm,y^\dn\}=2$). $\Phi$ and $\Psi$ are arbitrary
    fields and $\theta$ is a dimensionless constant. Note that, in new action, after above replacement all matrices in the YM theory
    have been changed with fields. Furthermore, these fields are functions of $x,y$ and hence
    $D_\dm X^i$ is no longer zero. Another point is that
    \eqref{prescription1} is right replacement when the number of
    D$_p$-branes are large.
    \item
    Fluctuations around two transverse directions, $X^\dm$, is
    considered as %
    \be\label{newx} %
    X^\dm\equiv\frac{1}{\lambda\theta}y^\dm-\epsilon^{\dm\dn}A_\dn.
    \ee %
    The form of above equation indicates that fluctuations play the
    role of extra components of the world volume D$_{p+2}$-brane gauge
    field which we will show it in detail.
\end{itemize}
Let's start with variation of scalar fields under gauge
transformation \eqref{xtransformatiom} which leads to %
\be\label{xtransformatiomnew} %
 \delta_gX^I=-(\delta_gy^{\dot{\mu}})\partial_{\dot{\mu}}X^I,
\ee %
where %
\be %
 \delta_gy^{\dot{\mu}}=\lambda\theta\epsilon^{\dot{\mu}\dot{\nu}}\partial_{\dot{\nu}}\alpha.
\ee %
Variation of the new components of the gauge field can be found by using
\eqref{newx} and \eqref{xtransformatiomnew} which is %
\be %
 \delta_gA_\dm=\partial_\dm\alpha+\lambda\theta\epsilon_{\dm\dn}\{X^\dn,\alpha\}.
\ee %

The first term in the YM action can be simplified by using
\eqref{newx} and we then have %
\be %
 (D_\mu X^I)^2=(D_\mu X^i)^2+(F_{\mu\dn})^2,
\ee %
The second term gives $\{X^i,X^j\}^2$ and %
\be\begin{split}\label{comrelation} %
 \{X^\dm,X^\dn\}^2&=\frac{2}{(\lambda\theta)^4}+\frac{1}{(\lambda\theta)^2}(F^{\dm\dn})^2+\ {\rm{total\ derivative\ terms}},\cr
 \{X^\dm,X^i\}^2&=\frac{1}{(\lambda\theta)^2}(D^\dm X^i)^2,
\end{split}\ee %
where %
\be\begin{split}\label{comrelation} %
 \{X^\dm,X^\dn\}&=\frac{\epsilon^{\dm\dn}}{(\lambda\theta)^2}-\frac{1}{\lambda\theta}\epsilon^{\dm\da}F_{\da\db}\epsilon^{\db\dn},\cr
 \{X^\dm,X^i\}&=\frac{\epsilon^{\dm\da}}{\lambda\theta}D_\da X^i ,
\end{split}\ee %
and %
\be\begin{split}%
 F_{\dm\nu}&=\partial_\dm A_\nu-\partial_\nu A_\dm-\lambda\theta\{A_\dm,A_\nu\},\cr
 F_{\dm\dn}&=\partial_\dm A_\dn-\partial_\dn A_\dm-\lambda\theta\{A_\dm,A_\dn\},\cr
 D_\dm\Phi&\equiv
 -\lambda\epsilon_{\dm\dn}\{X^\dn,\Phi\}=\partial_\dm\Phi-\lambda\theta\{A_\dm,\Phi\}.
\end{split}\ee %
By plugging the above expressions in \eqref{action}, we arrive at a
\emph{non-commutative} D$_{p+2}$-brane action is appeared %
\be %
 S=-\frac{1}{2g^2_{YM}\lambda}\int d^{p+3}x \left((D_{\hat{\mu}}
 X^i)^2+\frac{(\lambda \theta)^2}{4}\{X^i,X^j\}^2+\frac12 F_{\hat{\mu}\hat{\nu}}^2+\frac{1}{2(\lambda
 \theta)^2}\right),
\ee %
where %
\be\begin{split} %
 F_{\hat{\mu}\hat{\nu}}&=\partial_{\hat{\mu}} A_{\hat{\nu}}
 -\partial_{\hat{\nu}} A_{\hat{\mu}}+\lambda \theta\{A_{\hat{\mu}},A_{\hat{\nu}}\},\cr
 D_{\hat{\mu}}\Phi&=\partial_{\hat{\mu}}\Phi+\lambda
 \theta\{A_{\hat{\mu}},\Phi\}.
\end{split}\ee %
$\hat{\mu}=(\mu,\dm)$ denotes the world volume index of
D$_{p+2}$-brane and the above action is invariant under %
\be\begin{split} %
 \delta_gA_\mu&=\partial_\mu\alpha+\lambda \theta\{A_{\mu},\alpha\},\cr
 \delta_gA_\dm&=\partial_\dm\alpha+\lambda
 \theta\epsilon_{\dm\dn}\{X^\dn,\alpha\}.
\end{split}\ee %
As we will see in next section this action is a low energy limit of
\emph{non-commutative} D$_{p+2}$-brane action \eqref{nonaction}
where the B-field turns on as
$B_{\dm\dn}=\frac{1}{\theta}\epsilon_{\dm\dn}$. Hence the arbitrary
constant, $\theta$, plays the role of \emph{non-commutative
parameter} and it is acceptable to recover the commutative action
when $\theta\rightarrow0$. Obviously, the U(1) gauge theory action
for D$_{p+2}$-brane is recovered although there is an extra term
going to infinity in the action. This term can be considered as zero point energy
of multiple D$_p$-branes dissolved in the D$_{p+2}$-brane world
volume.

\section{ DBI theory}
One of the remarkable properties of D$_p$-brane is that the U(1)
gauge symmetry is enhanced to U(N) gauge symmetry for N coincident
D$_p$-branes. The suitable action for multiple
D$_p$-branes was introduced in \cite{Myers}. The action is %
\be\begin{split}\label{mdbiactin} %
 S&=-T_p\int d^{p+1}x\cr
 &\times{\rm{STr}}\Bigg(
 \sqrt{-\det\left({\rm{P}}[E_{mn}
 +E_{m I}(Q^{-1}-\delta)^{IJ}E_{Jn}]+\lambda
 F_{\mu\nu}\right)\det(Q^I_J)}\Bigg),
\end{split}\ee %
where $E_{mn}=G_{mn}+B_{mn},\
Q^I_J\equiv\delta^I_J+i\lambda[X^I,X^K]E_{KJ}$. The brane tension is
$T_p=\frac{2\pi}{g_s(2\pi l_s)^{p+1}}$($l_s$ and $g_s$ are string
length and coupling) and "P" denotes pull-back of background metric
and NSNS two-form($m,n=0,..,9$). $F_{\mu\nu}$ is field strength of
gauge field living on the D$_p$-brane. STr(...) denotes that
one takes a symmetrized average over all ordering of various fields appearing
inside the trace.

Let us again consider that two longitudinal directions of D-branes
fluctuate such that
$X^\dm=\frac{1}{\lambda\theta}y^\dm-\epsilon^{\dm\dn}A_\dn$.
Working, in static gauge \textit{i.e.} $\lambda X^{\mu}=x^{\mu}$, in
flat space background and noting that the $B$ is turned off in
D$_p$-brane theory, the determinant in the action can be
decomposed as
\be %
\tilde{D}=\det\left(%
\begin{array}{ccc}
  \eta_{\mu\nu}+\lambda F_{\mu\nu} & \lambda D_\mu X^\dn &  \lambda D_\mu X^j \\
  -\lambda D_\nu X^\dm & \delta^{\dm\dn}+i\lambda X^{\dm\dn} & i\lambda X^{\dm j} \\
  -\lambda D_\nu X^i & i\lambda X^{i\dn} & Q^i_k\delta^{kj} \\
\end{array}%
\right). %
\ee %
Using \eqref{comrelation} and
\be \label{fields}%
 D_{\mu}X^{\dn}=F_{\mu\da}\epsilon^{\da\dn},
\ee %
one can rewrite the determinant as
\be %
\tilde{D}=\det\left(%
\begin{array}{ccc} \label{nondbi}
  \eta_{\mu\nu}+\lambda F_{\mu\nu} & \lambda F_{\mu\da}\epsilon^{\da\dm} &  \lambda D_\mu X^i \\
  -\lambda F_{\nu\da}\epsilon^{\da\dm} & \delta^{\dm\dn}-\frac{1}{\theta}\epsilon^{\dm\dn}+\lambda \epsilon^{\dm\da}F_{\da\db}\epsilon^{\db\dn} & -\lambda \epsilon^{\dm\da}D_\da X^i \\
  -\lambda D_\nu X^i & \lambda \epsilon^{\dn\da}D_\da X^i & Q^i_k\delta^{kj} \\
\end{array}%
\right). %
\ee %
Using the standard trick for recombining the determinant\cite{Myers}
\be %
\tilde{D}=\det\left(%
\begin{array}{ccc} \label{nondbi}
  A & B \\
  C & D
\end{array}%
\right) %
=\det{(A-BD^{-1}C)\det{(D)}},
\ee%
one can rearrange the action in a compact form similar to
\eqref{mdbiactin}. As it can be seen, in the new action a
constant $B$-field is turned on in new directions $\dm$'s. Due to
the facts that the covariant derivatives $D_{\hat{\mu}}$ and
field strength $F_{\hat{\mu}\hat{\nu}}$ have now a non-vanishing first
order $\theta$ term and the $B$ field is present, the new action is the $\theta\rightarrow
0$ limit of a \emph{non-commutative} $U(1)$ D$_{p+2}$ brane action\cite{Ardalan}. 
The general form of such an action is
\be\begin{split}\label{nonaction} %
 \hat{S}&=-T_p\int d^{p+1}x\Bigg(e^{-\phi(\hat{A})}\cr
 &\times\sqrt{-\det\left({\rm{P}}_\theta[E_{mn}(\hat{A})
 +E_{m I}(\hat{A})(Q^{-1}-\delta)^{IJ}E_{Jn}(\hat{A})]+\lambda
 \hat{F}_{mn}\right)\det(Q^I_J)}\Bigg)*_{N}
\end{split}\ee %
where
$Q^I_J\equiv\delta^I_J+i\lambda[\hat{X}^I,\hat{X}^K]_ME_{KJ}(\hat{A})$ and "$\ \hat{}\ $"
denotes non-commutative fields. Here the $*_N$ is the multiplication rule
among various non-commutative fields \cite{Liu2}.
  One can show that in
the limits where the non-commutativity parameter $\theta$ goes to
zero this action reduces to DBI action.

\section{Non-BPS DBI action}
Besides of BPS branes, which are charges of RR fields, non-BPS branes may
exist in the IIA(B) theory. The existence of these branes may cause the instability
and breaks the supersymmetry. From the field theory of world volume of D-brane point of view,
the existence of a tachyonic field T is the reason of such phenomena.

In general, the action of non-BPS D$_p$-branes can be written as the
product of the action of BPS D$_p$-branes and contribution which comes from the tchyon field such as
\bea%
 S_{Non-BPS}=T_p\int{d^{p+1}x\ S_{BPS}F(T)},
\eea%
where $F(T)$ is the tachyonic contribution of the action of
D$_p$-brane.
The form of this action can be obtained by using the
non-BPS D$_9$(or D$_8$)-brane action and then performing the
T-duality transformations\cite{Kluson}. Its most general form up to
first order derivative of $T$ may be written as
\newpage
\bea%
F(T)=V(T)&-&\Sigma_{n=1}^{\infty}f_n(T)\lambda^n(( E^{\mu\nu}-E^{\mu
I}E_{IJ}E^{J\nu})D_{\mu}TD_{\nu}T\cr &+&2i E^{\mu
K}E_{KJ}[X^J,T]D_{\mu}T-E_{IJ}[X^I,T][X^J,T])^n,
\eea%
and $V(T)$ is the potential for the tachyon. Functions $f_n(T)$
are some even functions of the tachyon field \cite{Sen}, and although their explicit
form are not known, there
are some conjectures about it \cite{Garousi,Bergshoeff}.

Using the prescription \eqref{prescription1} in flat space
background the second term vanishes and  so $F(T)$ reads as
\bea%
F(T)=V(T)-\Sigma_{n=1}^{\infty}f_n(T)\lambda^n(( E^{\mu\nu}-E^{\mu
I}E_{IJ}E^{J\nu})D_{\mu}TD_{\nu}T\cr
+(\lambda\theta)^2E_{IJ}\{X^I,T\}\{X^J,T\})^n.
\eea%
Applying
$X^{\dm}=\frac{1}{\lambda\theta}y^{\dm}-\epsilon^{\dm\dn}A_{\dn}$
and noting that
 $$\epsilon_{\dm\dn}D^{\dm}TD^{\dn}T=0,$$ one easily finds that the
tachyonic part of Non-BPS multiple D$_p$-branes action changes to
tachyonic part of a Non-BPS D$_{p+2}$-brane action.

\section{Chern-Simons term}
In this section we examine that the Chern-Simons term of N D$_p$-brane action also
reproduce the Chern-Simons term of a single D$_{p+2}$-brane in the large N limits
discussed in previous sections. As it is well known, the world volume of D$_p$
branes in IIA(B) string theory can couple to RR fields of rank lower
than the dimension of brane and also can couple to RR fields with
higher rank due to Myers terms\cite{Myers}. In fact, the matrix representation
of fields and then the non-commutativity of such fields in non-Abelian theories
allows one to couple a combination of form fields of higher rank and commutators of
scalar fields with world volume of D-brane in a covariant manner.

On the other hand, the Chern-Simons action is given by \cite{Myers} %
\bea %
 S_{CS}=\mu_p\int{{\rm{STr}}\left({\rm{P}}[e^{i\lambda{\rm{i}_X}{\rm{i}_X}}
 (\Sigma C^{(n)}e^{B})]e^{\lambda F}\right),}
\eea %
where $C^{(n)}$ denotes $(p+1)$-forms and $\mu_p$ is the RR charge
of the brane and $\rm{i}_X$ is the exterior derivative in $X$ direction which acts on RR
form fields. This special form of the action is necessary for
compatibility with the T-duality transformations of various IIA and
IIB fields. Note that all fields are in adjoint representation of
the U(N) gauge group.

The general couplings of these form fields with world volume of
D$_p$-branes have many terms so, for simplicity, we study
D$_1$-D$_3$ transition and only consider the couplings of all form
fields up to the first power of fields.

Before we proceed, we mention that due to the
non-commutativity of two longitudinal directions of D$_1$-branes, one
may define \cite{Mukhi,Liu} a new two form field $Q^{(2)}$ which its
components are
\be %
Q^{(2)}\equiv-i\lambda^3\theta [X_{\dm},X_{\dn}]dX^{\dm}\wedge
dX^{\dn}.
\ee%
So, one should consider the couplings of this two form field with
world volume of D$_1$-branes. We will see that this form fields has
important role in this story.
Finally, we will do our computations in the regime where $\theta
\rightarrow 0$ and
also consider the constant $C_{\dm\dn}$ and $F_{\dm\dn}$ fields such
that
\be%
C_{\dm\dn}=\lambda^2\epsilon_{\dm\dn},\;\;\;\;\;F_{\dm\dn}=\frac{\epsilon_{\dm\dn}}{\lambda}.
\ee%
By the above assumptions we have
\bea%
\lambda^2\theta\{X^{\dm},X^{\dn}\}&\rightarrow&
\epsilon^{\dm\dn}(\frac{1}{\theta}-1),\;\;\;\;\;\;Q_{\dm\dn}\rightarrow
\lambda^2\epsilon_{\dm\dn},\cr
\lambda^2\theta\{X^{\dm},X^{i}\}&\rightarrow& \lambda
\epsilon^{\dm\da}D_{\da}X^{i},\;\;\;\;\;\;\theta\{X^i,X^j\}\rightarrow
0,
\eea%
After all, recalling that in D$_1$-branes we have $B=0$ then the
Chern-Simons term is equal to
\bea \label{csd1}%
 S_{CS}&=&\frac{1}{\lambda}\int{\rm{STr}}\left({\rm{P}}[e^{i\lambda \textbf{i}_X\textbf{i}_X}(
 C^{(2)}(1+Q^{(2)}))]e^{\lambda F}\right)\cr
  &\rightarrow &\frac{1}{\lambda^2}
  \int{d^4x\left(\frac{1}{2}C_{mn}D_{\mu} X^mD_{\nu} X^n\right)\epsilon^{\mu\nu}}\cr
  &+&\frac{1}{\lambda^2}\int{d^4x\left(\frac{-\lambda^2\theta}{2\lambda^2}\{X^{m},X^{n}\}C_{nm}\right)\lambda F_{\mu\nu}\epsilon^{\mu\nu}}\cr
  &+&\frac{1}{\lambda^2}\int{d^4x\left(\frac{-\lambda^2\theta}{2\lambda^2}\{X^{m},X^{n}\}C_{nm}\right)
  Q_{\dm\dn}D_{\mu} X^{\dm}D_{\nu} X^{\dn}\epsilon^{\mu\nu}}\cr
  &+&\frac{1}{\lambda^2}\int{d^4x\left(\frac{-\lambda^2\theta}{2\lambda^2}\{X^{\dm},X^{\dn}\}Q_{\dn\dm}\right)
  C_{mn}D_{\mu} X^{m}D_{\nu} X^{n}\epsilon^{\mu\nu}}\cr
  &+&\frac{1}{\lambda^2}\int{d^4x\left(\frac{-\lambda^2\theta}{2\lambda^2}\{X^{\dm},X^{m}\}C_{[m[n}Q_{\dm]\dn]}\right)
  D_{\mu} X^{n}D_{\nu} X^{\dn}\epsilon^{\mu\nu}},
\eea %
where the last term is antisymmetric for the pairs $(m,\dm)$ and
$(n,\dn)$. Note also that due to dimensions of form fields any
contraction with these fields leaves a factor $\frac{1}{\lambda}$ in
the action.

Using \eqref{comrelation} and \eqref{fields} one obtains
\bea%
C_{mn}D_{\mu} X^mD_{\nu} X^n&=&
\frac{1}{\lambda^2}C_{\mu\nu}-\frac{1}{\lambda}C_{i[\mu}D_{\nu]}X^i+C_{ij}D_{\mu}X^iD_{\nu}X^j\cr
&-&\frac{1}{\lambda}C_{\dm[\mu}F_{\nu]\da}\epsilon^{\da\dm} -C_{\dm
i}D_{[\mu}X^iF_{\nu]\da}\epsilon^{\da\dm}+{\cal{O}}(F^2),
\eea%
and one may rewrite the integrand of the above expression term by
term as
\bea%
&+&\frac{1}{2\lambda^2}\left(C_{mn}D_{\mu}
X^mD_{\nu}X^n\right)\epsilon^{\mu\nu}\cr
&-&\frac{1}{2\lambda^2}\left(2\lambda(\frac{1}{\theta}-1)+2C_{i\dm}D_{\da}X^i\epsilon^{\dm\da}+{\cal{O}}(\theta)\right)F_{\mu\nu}\epsilon^{\mu\nu}\cr
&-&\frac{1}{2\lambda^2}\left({\cal{O}}(F^2)\right)\cr
&-&\frac{1}{2\lambda^2}\left(2(\frac{1}{\theta}-1)C_{mn}D_{\mu}
X^mD_{\nu} X^n\right)\epsilon^{\mu\nu}\cr
&-&\frac{1}{2\lambda^2}\left(2(\frac{1}{\theta}-1)(C_{\mu\dn}+C_{i\dn}D_{\mu}X^i)F_{\nu\db}\epsilon^{\db\dn}+{\cal{O}}(F^2)\right)\epsilon^{\mu\nu}\cr
&-&\frac{1}{2\lambda^2}\left((C_{\mu
i}+C_{ji}D_{\mu}X^j)F_{\nu\db}D_{\dn}X^i\epsilon^{\db\dn}+{\cal{O}}(F^2)\right)\epsilon^{\mu\nu}
\eea%
where the two last terms come from the last term of \eqref{csd1}.

Now, for D$_3$-brane, the Chern-Simons term up to first power of
fields is equal to
\bea \label{chd3}%
 S_{CS}&=&\frac{1}{\lambda^2}\int{\rm{STr}}\left(P[e^{i \lambda \textbf{i}_X\textbf{i}_X}(\Sigma
 C^{(n)}e^B)]e^{\lambda F}\right)\cr
  &\rightarrow&\frac{1}{2\lambda^2}\left(\int d^2x\left(
  C_{mn}B_{pq}
  D_{\hat{\mu}} X^mD_{\hat{\nu}} X^nD_{\hat{\lambda}} X^p D_{\hat{\theta}}
  X^q\right)\epsilon^{\hat{\mu}\hat{\nu}\hat{\lambda}\hat{\theta}}+\rm{antisymm\;part}\right)\cr
  &+&\frac{1}{2\lambda^2}\left(\int{d^2x\left(C_{mn}D_{\hat{\mu}}
  X^mD_{\hat{\nu}} X^n\lambda F_{\hat{\lambda}\hat{\theta}}\right)
  \epsilon^{\hat{\mu}\hat{\nu}\hat{\lambda}\hat{\theta}}}+\rm{antisymm\;part}\right),
\eea %
where by \textit{antisymm part} we mean that, for example, the terms
of $C_{mn}B_{pq}$ are antisymmetric under the exchange of indices of
$C$ and $B$ fields. Noting that in D$_3$-brane we have found
$P(B)=\frac{\epsilon_{\dm\dn}}{\theta}$, we rewrite the $S_{CS}$
of D$_3$ brane as\newpage
\bea %
S_{CH}&=&\frac{1}{2\lambda^2}\int{d^2x
 \left(C_{\mu\nu}-2C_{i\mu}D_{\nu}X^i+C_{ij}D_{\mu}X^id_{\nu}X^j\right)\epsilon_{\dm\dn}\epsilon^{\mu\nu\dm\dn}}\cr
 &+&\frac{1}{2\lambda}\int{d^2x\left(C_{mn}D_{\hat{\mu}}X^mD_{\hat{\nu}}X^{n}F_{\hat{\lambda}\hat{\theta}}\right)
 \epsilon^{\hat{\mu}\hat{\nu}\hat{\lambda}\hat{\theta}}}.
\eea %
It is not hard to show that the above D$_3$-brane Chern-Simons terms
reproduce the D$_1$-branes Chern-Simons terms. We see the role of
the two form field $Q^{(2)}$ in this D$_1$-D$_3$ transition in which
some terms in D$_3$ such as $C_{\mu
i}F_{\nu\db}D_{\dn}X^i\epsilon^{\db\dn}\epsilon^{\mu\nu}$ come from
the coupling of this form field with $C^{(2)}$ and $F^{(2)}$ in
\eqref{csd1}.

The general form of Chern-Simons action in non-commutative theories
in the presence of
non-zero constant B field may be written as\cite{Mukhi,Liu,Mukhi2}%
\be %
 S_{CS}=\mu_p\int\frac{\rm{Pf}\hat{Q}}{\rm{Pf}\theta}\left(P[e^{i(i_X*i_X)}(\Sigma
 C^{(n)})]e^{B+\lambda F}\right), %
\ee %
where
$\hat{Q}^{mn}=\theta^{mn}-\theta^{m\hat{\alpha}}\hat{F}_{\hat{\alpha}\hat{\beta}}\theta^{\hat{\beta}n}$
and $\theta^{mn}=(B^{-1})^{mn}$, the Pf denotes the Pfaffian of an antisymmetric matrix and
all products are understood as $\ast$ product. It can be seen that in the $\theta\rightarrow 0$ limit
this action and the action \eqref{chd3} coincide with each other.

\section{Conclusion}
In this paper we have shown that in flat space multiple
D$_p$-branes can be considered as a D$_{p+2}$-brane for large number
of  D$_p$-branes. In fact matrix valued $p+1$ dimension fields go
to $p+3$ dimension fields. The point is that two transverse
directions to multiple D$_p$-branes appear as two components of gauge
field living on the D$_{p+2}$-brane. One can run this prescription
in opposite way and start with a D$_{p+2}$-brane and
it finally leads to multiple D$_p$-branes. In other words, one exchanges two
scalar fields with two components of gauge fields and at the end we
have a D$_{p+2}$-brane or multiple D$_{p-2}$-branes.

Although, this setup has been done in flat space the above idea
can be generalized to curved background. In well-known curved background such as
pp-wave our half BPS D$_p$-branes have spherical symmetry and we may
then expect multiple D$_p$-branes lead to a spherical D$_{p+2}$-brane.

Moreover, by using the above prescription, we expect a relation
between mN D$_{p}$-branes and m D$_{p+2}$-branes in large N limit. Such idea
is useful to extract new understanding of the theory of multiple non-commutative D-branes.

Understanding of three dimensional conformal field theory(CFT) was an open
issue for years \cite{Schwarz}. Symmetries of this theory are consistent with
symmetries of multiple M2-branes. Recently, a groundbreaking three dimensional CFT was presented in
\cite{Bagger} known as BLG theory.
In BLG theory, there are a lot of attempts to show that a M5-brane action
leads to multiple M2-branes action or vice versa. We hope that uplifting the above
results teach us more about M2-M5 relation. In the other hand in comparison to
non-commutative D$_p$-brane, it seems that one should know about "\emph{non-commutative}"
M5-brane to explain the relation correctly, although the geometry of M5-brane
is not known by now \cite{Chen}.

\section*{ Acknowledgment}
We would like to specially thank M. M. Sheikh-Jabbari for careful
reading of the manuscript and valuable discussion.

\end{document}